\documentclass[angeod,manuscript]{copernicus}  

\nolinenumbers
\usepackage{amsmath}
\usepackage{color}

\usepackage{natbib}

\begin{document}

\title{{ECMI Resonance in AKR Revisited:  Hyperbolic Resonance, Harmonics, Wave-Wave Interaction 
}}

\author[1]{W. Baumjohann}
\author[2]{R. A. Treumann}
\affil[1]{Space Research Institute, Austrian Academy of Sciences, Graz, Austria}
\affil[2]{International Space Science Institute, Bern, Switzerland}
\affil[3]{Institute of Geophysics, Munich University, Germany

\emph{Correspondence to}: Wolfgang.Baumjohann@oeaw.ac.at
}

\runningtitle{AKR Resonance}

\runningauthor{Baumjohann \& Treumann}

\received{ }
\pubdiscuss{ } 
\revised{ }
\accepted{ }
\published{ }


\firstpage{1}

\maketitle

\begin{abstract}
Recapitulation of the resonance condition for the fundamental and higher electron cyclotron harmonics in the Electron Cyclotron Maser Instability (ECMI) enables radiation below and confirms the possibility of radiation in a narrow band above harmonics $n>1$. Near $n=1$ resonance on the confined lower X-mode branch, amplification is supported by the decrease of phase and group speeds. Confined slow large-amplitude quasi-electrostatic X-modes nonlinearly modulate the plasma to form cavitons until self-trapped inside them at further increasing wavenumber. They undergo wave-wave interaction, enabling escape to free space in the second harmonic band below $n=2$. At sufficiently large parallel wavenumber (oblique propagation), the fundamental resonance $n=1$ is hyperbolic, a possibility so far missed but vital for an effective ECMI in the upward current region. Here, the resonance hyperbola favourably fits the loss cone boundary, the presumably important ECMI upward-current source-electron distribution, to  stimulate ECMI growth at available auroral electron energies.    

\end{abstract}
\section{Introduction}
   
The Electron Cyclotron Maser (ECMI) instability mechanism \citep{twiss1958,hirshfield1977,melrose1986,treumann2006} has, for already close to half a century  \citep[since][]{wu1979}, been identified as the canonical mechanism of generating AKR, the celebrated ``auroral kilometric radiation'' \citep[first analyzed by][who identified its auroral origin]{gurnett1974}, a sporadic narrow-banded intense radio emission \citep[discovered by][in hardly accessible work]{benediktov1965}. It is emitted preferentially during substorms, the major disturbance \citep{kamide1993,akasofu2004,akasofu2021} of earth's magnetosphere. 

AKR \citep[for a recent review of observations, cf., e.g.,][]{yearby2022} propagates in the X-mode polarization and is radiated from the auroral zone close to earth where the magnetic field is strong and the thermal plasma on various scales becomes locally diluted, frequently with very small plasma-to-cyclotron frequency ratio $\omega_e/\omega_{ce}\ll1$. Under such circumstances the ECMI sets on if only its \emph{necessary} and \emph{sufficient} conditions are both  met simultaneously. These refer to the cyclotron resonance of weakly relativistic electrons occupying an excited energy state \citep[first suggested in more general astrophysical context by][]{twiss1958} and, for the sufficient condition, to a suitable electronic energy source. In classical physics an excited non-thermal state is identified with some highly non-thermal electron phase-space distribution function that deviates strongly from a thermal \citep[even anisotropic relativistic, see][]{treumann2016} Boltzmann-Maxwell distribution. Experimentally, the relevant distribution that would be effective in exciting the ECMI has so far not yet been ultimately identified. It is conventionally assumed to be kind of a simple (weakly relativistic) loss-cone distribution \citep[cf., e.g.,][and many others]{wu1979,pritchett1984} in phase space as has frequently been measured \citep[for a typical measurement in situ the auroral magnetosphere, cf., e.g.][]{delory1998} in the presumable  AKR source region, an assumption which has however been challenged \citep{labelle2002,baumjohann2022}. 

In a recent note \citep{baumjohann2022} we inspected the paradigmatic mechanism of auroral kilometric radiation, the most intense natural radio emission in near-earth space. We noted a number of  problems with that mechanism which have not been or were at most approximately solved yet. Nevertheless it seems to be certain that the ECMI is the sole sufficiently strong, and thus reasonable and probable, mechanism that could explain the emission of the surprisingly intense radiation pattern of AKR. Simultaneously it may also serve as the non-thermal radio emission paradigm in strongly magnetized objects for application in other places in space: planetary magnetospheres in general, extrasolar planetary systems, and probably as well in astrophysics to generate sporadic intense emissions in the radio band which cannot be understood as synchrotron radiation but must have been caused by unknown non-thermal electron distributions, even though the most effective of these distributions still have to be identified unambiguously. 

In the present note we do not go into the important question of the sufficient condition: What kind of distributions are responsible for ECMI. We rather deal merely with the necessary condition that has to be clarified first, prior to worry about which most probable and experimentally supported phase-space distribution must be chosen to obtain sufficient radiation intensities. This is the question we attack here to be answered: Given \emph{any suitable} distribution function, under which conditions can the ECMI become excited  and escape into free space, hopefully at the observed large amplitudes? This question is related to kinetic theory and the condition of resonance between the ECMI-source electrons and the relevant electromagnetic wave mode.

\section{Resonance -- the necessary condition}
 
The ECMI is a kinetic instability emitted in the X mode polarization \citep[cf., eg.,][]{melrose1986,treumann2006}. By reference to observations it propagates (predominantly) in the electromagnetic X-R mode \citep{krall1973,baumjohann1996} almost perpendicular to the strong ambient magnetic field $\vec{B}$.  In the ECMI the X mode is excited when, as noted above, the electron component in the plasma exhibits a particular momentum space distribution function, representing the equivalent of an elevated energy state of the resonant electron component. 

Assume a relativistic electron velocity (momentum) distribution $f_e(\vec{p})$ of such an energetically elevated kind. In a certain range of frequencies $\omega$ and wave numbers $\vec{k}$, it obeys the required property and allows for the excitation of the ECMI with growth rate 
\begin{equation}\label{eq-growthrate}
\Gamma(\omega,\vec{k})=-\mathrm{Im}\,\mathcal{D}(\omega,{\vec{k}})/\partial_{\omega}\mathrm{Re}\,\mathcal{D}(\omega,\vec{k})\big|_{\omega_X}=\sum_n\Gamma_n(\omega,\vec{k})>0. 
\end{equation}
where $\omega(\vec{k})=\omega_X(\vec{k})$ is the X-mode frequency as function of wave number $\vec{k}$, and $\mathcal{D}(\omega,\vec{k})$ is the kinetic dispersion relation, explicitly given in \citep{krall1973,baumjohann1996} and, for a loss-cone distribution, explicated in \citep{wu1983}, taken at the X-mode frequency and wave number \citep[for the full magneto-ionic theory\footnote{It should be noted that the magneto-ionic theory is a fluid theory. It assumes that those wave modes are present, while the hot kinetic plasma component just serves its excitation or damping. In linear dispersion this permits neglecting all nonlinear modifications.} of the X mode in a cold electron-proton plasma cf., e.g.,][]{budden1988,melrose1986}. $\mathcal{D}(\omega,\vec{k})$  results from phase space integration over the source electron distribution function $f_e(\vec{p})$ in a homogeneous magnetized plasma and its gyrotropic derivatives with respect to the components $p_\|,p_\perp$ of $\vec{p}$, a quite involved expression in a non-Maxwellian plasma even for purely perpendicular wave propagation of the X mode. 

In instability each resonance $n$ provides a positive/negative contribution to the total positive growth rate $\Gamma(\omega,\vec{k})>0$ at the particular frequency. This growth rate, being a function of the X mode frequency $\omega_X$, accordingly exhibits discrete maxima/minima at or near to each resonance $n$, depending on the contribution of $\Gamma_n(\omega,\vec{k})$ being positive or not at that particular frequency. Plasmas like the upper-auroral substorm-magnetosphere that are capable of exciting the ECMI with growth rate $\Gamma>0$, are dilute with $\omega_{ce}\gg\omega_e$ and, for the purposes of X-mode propagation can, sufficiently justified, be considered cold and non-relativistic, as it is only the weakly-relativistic non-thermal auroral electron population, being of different than ionospheric origin and responsible for the unstable excitation of the ECMI. This may be different in much hotter astrophysical plasmas, however, where the relativistic modification in the X-R mode dispersion may have to be taken into account, a problem we are not concerned about here in the context of the magnetosphere. The assumed low density state relegates the source  of AKR and excitation of ECMI to the spatially extended auroral upward current region and generally excludes the narrow and dense downward current region, which also implies that the prevalent electron distribution is believed to be of the loss-cone family. This notion has been challenged \citep{baumjohann2022,treumann2012} though not disproved. In the following no attempt is made to calculate the growth rate for any specific distribution, which leaves this question open. 

\subsection{Harmonic resonances $n>1$}
The necessary condition for contributing to instability at harmonic number $n>1$ and parallel or antiparallel propagation
\begin{equation}\label{eq-n-res}
\omega-n\omega_{ce}/\gamma - k_\|c\beta_\| =0
\end{equation}
with $k_\|=k\cos\theta$ is the $n$th resonance in the phase space integral, with $n= \pm1, \pm2, \dots$, the cyclotron harmonic number which, in calculating the total growth rate $\Gamma(\omega,\vec{k})$ as function of the X mode frequency $\omega=\omega_X$, is summed over. Because the relativistic factor $\gamma>1$, their fundamental (relativistic) resonant frequency $\omega_{1X} =\omega_{ce}/\gamma$ at $n=1$ is always below the non-relativistic cyclotron frequency $\omega_{ce}=eB/m_e$ and thus confined to the plasma while in all higher harmonics $|n|>1$ the resonant frequencies $\omega_{nX}$  exceed the X mode stop band $\omega_{uh}\leq\omega_X\leq\omega_{co}^u$ between the upper hybrid $\omega_{uh}=\omega_{ce}\sqrt{1+\omega_e^2/\omega_{ce}^2}$ and upper cut-off $\omega_{co}^u=\frac{1}{2}\omega_{ce}(\sqrt{1+4\omega_e^2/\omega_{ce}^2}+1)$ frequencies.  While the fundamental cannot escape without help, the higher harmonics would naturally radiate away at about the speed of light if excited, having had little time for amplification and thus should generally be rather weak. 
 
 In the growth rate sum, each term $\Gamma_n(\omega,\vec{k})$ gives the contribution of the harmonic resonance $n$ to $\Gamma(\omega,\vec{k})$ or, in other words, the growth rate of the harmonic frequency at number $n$. The harmonic number $-\infty<n<\infty$  introduces an infinite chain of possible resonances from which the given distribution function $f_e(\vec{p})$ and the available resonant electron energies $\epsilon_e=m_ec^2\gamma(\vec p)$ select. For emission, the AKR source must be relativistic \citep{wu1979,melrose1986,melrose1994,treumann1997},  which is taken care of in $\gamma(\vec{p})=\sqrt{1+p^2/m_e^2c^2}$, and the electron momentum $\vec{p}=m_e\gamma c\vec{\beta}$, where $\vec{\beta}\equiv \vec{v}/c$ is defined.    

Working in momentum space $\vec{p}$ is more convenient. It has the advantage that  no restriction is to be made on $\vec{p}$ as the fully relativistic momentum spreads the entire space $0\leq |p|<\infty$ thereby avoiding the inconvenient upper bound on $\beta\leq1$, the circle of light speed radius in $\vec\beta$-space which apparently excludes any high-speed beams for becoming involved. This artefact disappears in momentum space.

For any $\gamma\gtrsim1$ and $\kappa_\|=k_\|c/\omega_{ce}=kc\cos\theta/\omega_{ce}$ the fully relativistic resonance condition reads
\begin{equation}\label{eq-beta-res}
\gamma(\vec{p})x-\kappa_\|p_\|/m_ec= n
\end{equation}
where $x=\omega/\omega_{ce}$ is the normalized frequency. Introducing the parallel momentum shift $p_0=nm_ec\kappa_\|/(x^2-\kappa_\|^2)$ yields the resonance line in gyrotropic momentum space $(p_\|,p_\perp)$ which is an ellipse
\begin{equation}\label{eq-p-res1}
\frac{(p_\|- p_0)^2}{a_{\|p}^2}+\frac{p_\perp^2}{a_{\perp p}^2}=1,\quad\mathrm{with}\quad 0<p^2<\infty
\end{equation}
where the two radii are given by
\begin{eqnarray}\label{eq-p-res2}
a_{\|p}^2&=&m^2_ec^2x^2(n^2+\kappa_\|^2-x^2)/(x^2-\kappa^2_\|)^2\\
a^2_{\perp p}&=&m^2_ec^2(n^2+\kappa^2_\|-x^2)/(x^2-\kappa_\|^2)
\end{eqnarray}
For $\kappa_\|=k_\|=0$ the ellipse degenerates into an unshifted circle, and all perpendicular harmonic emissions are below $x<n$ for any energy $\gamma$. Usually the resonance ellipse is in the literature expressed in $\vec{\beta}$. There is a subtle difference between the two representations of $\vec{\beta}$ and $\vec{p}$ resonances. The former is insensitive to the relation between frequency $x$ and parallel wavenumber $\kappa_\|$, whereas in the latter this relation may affect the shape of the resonance curve, a problem we will return to later. 

\begin{figure}[t!]
\centerline{\includegraphics[width=0.5\textwidth,clip=]{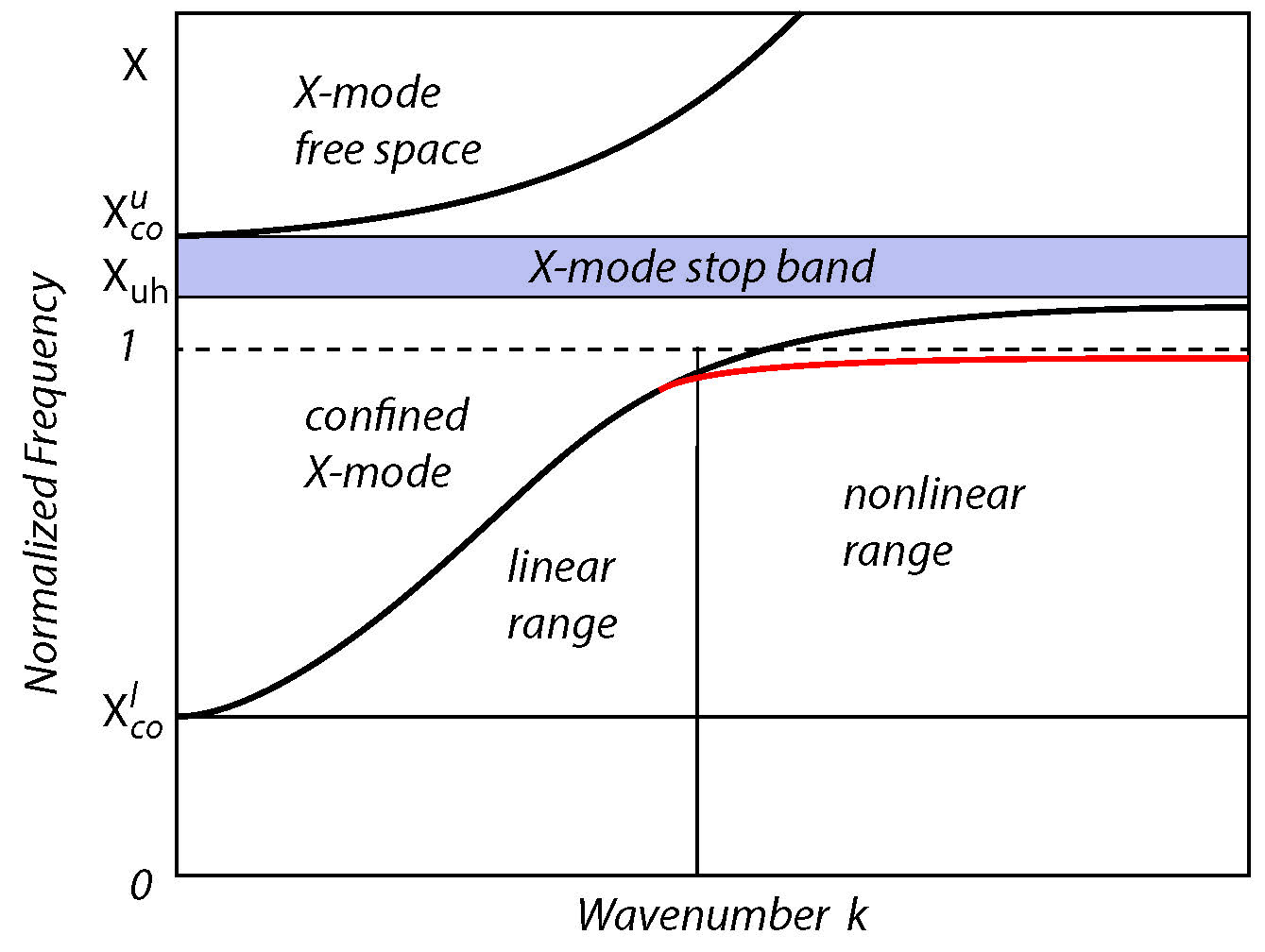}}
\caption{The plasma-confined X-mode dispersion relation in the auroral upward current AKR source region (lower solid curve) under the prevalent low-density plasma condition $\omega_e\ll \omega_{ce}$. The upper curve is the free-space X-mode. In the stop band between both X-mode branches the X mode cannot propagate. Normalization of the frequencies is to $x\equiv\omega/\omega_{ce}$, with $x=1$ the non-relativistic electron cyclotron frequency which is crossed by the linear lower-branch dispersion curve slightly above the relativistically allowed ECMI resonance. It propagates between the lower cut-off and upper hybrid frequencies. The red line approaching $x=1$ is the \emph{nonlinear} X-mode dispersion branch at maintained resonant frequency $x_\mathit{res}\approx\mathrm{const}\lesssim1$, stretching out to large wave-numbers $k$. Note that here on the dotted line the linear dispersion (\ref{eq-disp-rel}) is invalid. In this range both the phase and group velocities of the lower-branch X mode at resonant frequency, which is held constant remaining below $\omega_{ce}$, are substantially reduced, with the wave becoming quasi-electrostatic in resonance with an appropriate distribution of electrons. Wave-wave interaction becomes probable here and results in escaping radiation at $x\lesssim 2$ on the free space upper X-mode branch. For 4-wave interaction also higher harmonics may become excited. No deformation of the dispersion relation as function of propagation angle $\theta$ is shown here, which would slightly modify resonance, dispersion and interaction and under large wavenumber conditions even switch from elliptic to hyperbolic (for the hyperbolic case see below).
} \label{fig-disprel}
\end{figure}

In oblique propagation the case $x^2>n^2+\kappa_\|^2$ is excluded, as both radii would become imaginary. In order to  maintain an ellipse, one requires that $a_{\perp p}^2>0$ is positive, which implies that $x>\kappa_\|$ and $n^2>x^2-\kappa_\|^2$. Of course, emission is  below 
\begin{equation}
x<\sqrt{n^2+\kappa_\|^2} 
\end{equation}
The higher the harmonic number, the more oblique is the wave mode which contributes. For small obliqueness $\theta\lesssim\pi/2$, the most interesting case in the free space ECMI with $\omega_{Xn}=kc$, this sets a condition on the angle of emission for extension above harmonic $n$ in the frequency band
\begin{equation}
n<\frac{\omega_{Xn}}{\omega_{ce}}<\frac{n}{|\sin\theta|}
\end{equation}
which is a narrow range only. There is no restriction on any emission below $n$, other than that the frequency must exceed the upper frequency cut-off $\omega_{co}^u$. This gives for small ratio $\omega_e/\omega_{ce}<1$ the auxiliary condition
\begin{equation}
1+\frac{\omega_{e}^2}{\omega_{ce}^2}<\frac{\omega_{Xn}}{\omega_{ce}}=\frac{n}{\gamma}
\end{equation}
for excitation and propagation in the $n$th harmonic. An upper limit $\gamma\lesssim n$ is set on the electron energy $\epsilon_e=m_ec^2\gamma$ in resonance with the $n$th harmonic to contribute to the radiation near the upper cut-off, meaning that all resonances contribute to radiation in the lowest band of the free-space X mode which, therefore, will be the most intense.
 
Below $n$ resonances and emissions are, in principle, unrestricted. However, the last condition also allows just for a narrow range $|n|<x<\sqrt{2}\ |n|$ of frequencies above $|n|$, depending on the electron distribution function and resonant energy $\gamma$ whether or not it favours ECMI. At higher frequencies above this limit the resonance condition does not permit  radiation, a general necessary condition which is independent of any particle distribution function. It weakly contradicts the claims in \citep{wu1983} while at the same time making them precise. Generally harmonic emissions will always be below the harmonic $n$ with just that narrow range of oblique emission above $n$. Observation of high-frequency bounded radiation can be taken as measurement of $n$.

However, though harmonic resonances (and in case of causing instability realized as radiation) are possible, they  necessarily will be rather weak because the radiation is on the upper X mode branch where it readily escapes from the source region at light velocity $c$. The most intense directly excited emission results from below the second harmonic $x=\omega/\omega_{ce}< 2$, just above the upper frequency cut-off  
\begin{equation}
\omega_{co}^u=\textstyle{\frac{1}{2}}\omega_{ce}\Big[\Big(1+4\omega_e^2/\omega_{ce}^2\Big)^{1/2} + 1\Big] < \omega_{nX}
\end{equation}  
where also all higher harmonics contribute if present. In the magnetosphere only radiation below $n=2$ is of interest, being restricted to frequencies $\omega_{co}^u<\omega_{2X}<2\omega_{ce}$ and yielding for the resonant electron energy the upper limit 
\begin{equation}
1< \gamma<2/\big(1+\omega_e^2/\omega_{ce}^2\big)
\end{equation}
This implies that in the auroral AKR source region electrons of energy $\epsilon_e=m_e\gamma c^2 \lesssim 0.9$ MeV are eligible of directly exciting AKR beneath the second X-mode harmonic in about perpendicular propagation in free space, if only their distribution function suites the ECMI instability. This range of necessary electron energies indeed includes auroral energies suggesting that second harmonic AKR radiation can naturally originate from the auroral magnetosphere, as was proposed by \citet{wu1983} and suggested from observation \citep{pottelette2005}. Radiation should however be weak for the above mentioned reason of unrestricted escape and lack of spatial amplification. Generally auroral electron energies are far below this limit \citep[however, for a different observation cf.,][]{xiao2016}. They barely are capable of exciting higher harmonics. For this simple reason, harmonics $n>2$ should be absent in AKR.

\subsection{Ultra-relativistic resonance}
The absence of any restriction on the momentum $\vec{p}$  suggests that even very high resonant electron momenta/energies could in general participate in the ECMI. Turning to the extreme, this raises interest in the ultra-relativistic limit, which here is listed for completeness, not for application in the magnetosphere but probably be of vital importance in astrophysics. 

Ultra-relativistic conditions imply $p^2\gg m_e^2c^2$ and  $\gamma(\vec{p})=p/m_ec$. The resonance ellipse has radii
\begin{equation}
\frac{a^2_{\|p}}{m_e^2c^2}=\frac{n^2x^2}{(x^2-\kappa_\|^2)^2}, \quad 
\frac{a^2_{\perp p}}{m_e^2c^2}=\frac{n^2}{x^2-\kappa_\|^2}
\end{equation}
It degenerates into a circle in perpendicular propagation $\kappa_\|=0$ of radius $a_p=nm_ec/x$. Generally one has
\begin{equation}
\frac{\omega}{\omega_{ce}}=\frac{n}{p/m_ec}+\kappa_\|\cos\theta_p
\end{equation}
where $\theta_p$ is the angle of the momenta of the resonant particles. The bounds on either $n$ or $p$ are obtained when replacing the frequency with the upper X-mode cut-off
\begin{equation}
\frac{|n|}{p/m_ec}>\big|1+\frac{\omega^2_e}{\omega^2_{ce}}-\kappa_\|\cos\theta_p\big|,\quad |n|>1
\end{equation}
Remember that $n$ can be both positive and negative, as it enters the sum of harmonic growth rates $\Gamma_n$. The second condition on $n$ just points out that this limitation applies to all harmonic emissions propagating in the free space X-R mode.

\subsection{Resonance at the fundamental}
In view of application to the terrestrial magnetosphere it has been correctly claimed \citep{wu1979} that the (weakly relativistic) ECMI resonance dominates at harmonic number $n=1$, the fundamental. The reasons for this we have already listed: Any higher harmonic has barely time for being amplified substantially before leaving the source. Therefore the apparent theoretical \citep{wu1983} and suspected or real observationally claimed presence of any intense second harmonic  AKR emission in the magnetosphere can hardly be understood as direct excitation. 

Restricting to resonance on the lower X-mode branch eliminates all harmonics $n>1$. With $0\leq\gamma-1=\sqrt{1+p^2/m^2c^2}-1\ll1$ this is true in the auroral region, from where the ECMI wave cannot escape into free space (Fig. \ref{fig-disprel}), as this requires traversing the X-mode stop band $\omega_{uh}\leq\omega\leq\omega_{co}^u$ between the upper hybrid and upper cut-off frequencies. 

Resonance is generally not restricted to proximity to $\omega_{ce}$. The lower X-mode branch for zero wave number $k\approx0$ starts at the lower cut-off frequency $\omega_X>\omega_{co}^l=\frac{1}{2}[(\omega_{ce}^2+4\omega_e^2)^{1/2}-\omega_{ce}]\approx \omega_e^2/\omega_{ce}$ for small ratios of plasma-to-cyclotron frequency, the parameter range where the ECMI becomes effective. Assuming strictly perpendicular propagation $k_\|=0$   all relativistic factors
\begin{equation}\label{eq-gamma-1}
1<\gamma<\omega_{ce}^2/\omega_e^2
\end{equation}
of electrons in the distribution function could in principle participate in the resonance. The oblique relativistic resonance condition for those electrons is given by (\ref{eq-beta-res}) with $n=1$. The parallel shift (now conventionally expressed in $\vec{\beta}$) becomes $\beta_0= \kappa_\|/(1+\kappa_\|^2)<1$, and the elliptic radii are
\begin{equation}
\frac{a_\|^2}{m_e^2c^2}=\frac{1+\kappa^2_\|-x^2}{(1+\kappa_\|^2)^2},\quad \frac{a_\perp^2}{m_e^2c^2}=1-\frac{x^2}{1+\kappa_\|^2}
\end{equation}

With large $\gamma\lesssim\omega^2_{ce}/\omega^2_e$ the resonance is placed far below $\omega_{ce}$ on the lower X-mode branch by the factor $\omega_e/\omega_{ce}$ in frequency, if only sufficiently high energy electrons would be available, implying that the resonant energy in this case would be high. In the magnetosphere we have $\omega_{ce}^2/\omega_e^2\gtrsim 10^{2}$ or even larger which, well away from the cyclotron frequency, yields  $\epsilon_e=m_e\gamma c^2 \sim 50$ MeV. Such (about ultra-relativistic) electrons are absent there which, in retrospect, justifies the assumption of proximity of the resonance to $\omega_{ce}$. \citep[We note, however, that high energies have been referred to by][for apparently observed (A)KR whose source the authors attribute to the radiation belts.]{xiao2016} In remote astrophysical systems, on the other hand, containing very-high, mostly even ultra-relativistic electrons, proximity to the cyclotron frequency may become spurious, and excitation of lower branch trapped long wavelength $k_\perp\gtrsim0$ X-mode waves near the lower cut-off  $\omega^l_{co}<\omega_X^l\ll\omega_{ce}$ may become possible or should be the rule. There the ECMI would include  almost the entire lower X-mode branch. 

A broad energetic electron distribution may undergo resonance over a large part of the lower X-mode branch contributing to the ECMI and growth of lower branch X-mode waves in a wide range of frequencies $\omega_X^l<\omega_X(k)<\omega_{ce}$. The wave numbers in resonance are obtained from the pure lower branch electron X-mode dispersion relation in perpendicular propagation \citep{krall1973,baumjohann1996}
\begin{equation}\label{eq-disp-rel}
\frac{k^2c^2}{\omega_{ce}^2}=\frac{(\omega_{co}^{u2}-\omega^2)(\omega^2-\omega_{co}^{l2})}{\omega_{ce}^2(\omega_{uh}^2-\omega^2)}
\end{equation}
where for our purely electronic purposes the ion contributions as well as higher powers of small quantities have been neglected. 

This dispersion curve crosses the cyclotron frequency at $\omega=\omega_{ce}$, corresponding to $k_{m\perp}\approx \sqrt{2}\ \omega_e/c$ which is about the inverse electron inertial length $\lambda_e=c/\omega_e$, setting an upper limit on the resonant wave number $k_\perp$, a large number respectively short wavelength indeed.  The unstable range $\Delta k_\perp$ maps to the energy range $\Delta \gamma(k_\perp)$ of resonant electrons 
\begin{equation}
0< k_\perp (\omega,\gamma)\ \lesssim \sqrt{2}/\lambda_e
\end{equation}
 Depending on the electron distribution function, instability becomes possible almost everywhere along the lower X mode branch if only sufficiently large electron energies are available. $k_\perp=0$ corresponds to reflection at the lower X-mode cut-off $\omega_{co}^l$ and is clearly outside the unstable wavenumber range.

Near $\omega\approx\omega_{ce}$ the phase velocity of the X mode becomes $v_X\approx c/\sqrt{2}$ implying that at an observed frequency $\omega\lesssim\omega_{ce}\approx 300$ kHz, the nominal magnetospheric AKR frequency \citep{gurnett1974}, the phase speed is low enough for the wave to experience several amplifications before leaving the source. 

In order to check this, we obtain for the group velocity
\begin{equation}
v_{gX}=\frac{d\omega}{dk_\perp}\approx -2^{3/2}\Big(\frac{\omega_e}{\omega_{ce}}\Big)^2c \ll -c
\end{equation}
which is much smaller than the velocity of light. For an ECMI growth rate $\Gamma(\omega_X)$ the spatial amplification rate $K\approx - \Gamma/v_{gX}$  becomes 
\begin{equation}
K(\omega) \approx \frac{\Gamma(\omega)}{2\sqrt{2}c}\Big(\frac{\omega_{ce}}{\omega_e}\Big)^2
\end{equation}
a comparably large value. Assume $\omega_{e}/\omega_{ce}\lesssim0.1$, as approximately applicable to the auroral source region, the group velocity is just the order of $v_{gX}\lesssim 10^3$ km/s, in the auroral magnetosphere this compares to the Alfv\'en velocity $v_A$, giving roughly $K\gtrsim 2\times10^{-4} \Gamma(\omega_X)$ km$^{-1}$.  So, for a growth rate $\Gamma\sim 10^{-4}\omega$, we have at the nominal emission frequency of $\omega\sim300$ kHz that 
\begin{equation}
K\gtrsim 4\times10^{-1}~\mathrm{km}^{-1}
\end{equation}
This corresponds to several e-folding lengths over one wavelength, sufficient for excitation of moderate (or even large) amplitude X-mode waves. Of course, growth rates $\Gamma(\omega)$ of this order may be extreme. Reducing $\Gamma(\omega)$ by two orders of magnitude still gives one e-folding over one wavelength of $\lambda_X\sim 1$ km within the oscillation and growth times. Such amplification rates are still large. 

This estimate suggests that the ECMI on the lower X-mode branch is  capable of generating rather large amplitude confined X-mode waves close to the local non-relativistic cyclotron frequency $\omega_{ce}$ predominantly because there the group velocity is strongly reduced, and the wave has sufficient time to grow before leaving the source. 

Growth becomes substantial here if the source is sufficiently extended over a number of wavelengths in the direction of propagation about perpendicular to the magnetic field $\vec{B}_0$. This very fact makes the ECMI on the lower X-mode branch (see Figure \ref{fig-disprel}) very interesting not just by itself but also for its possible nonlinear interaction with the dilute plasma component whose pressure is necessarily low and thus susceptible to large amplitude electromagnetic radiation and the implications of its pondero-motive force. Modulation instability of the unstable ECMI may become possible causing chains of solitons and structure in the plasma, effects which have not yet been considered in connection with ECMI but may play an important role if investigating its fine structure \citep[compare Figure 2 in][]{baumjohann2022} and the reaction/response of the auroral environment to its presence.

\begin{figure}[t!]
\centerline{\includegraphics[width=0.75\textwidth,clip=]{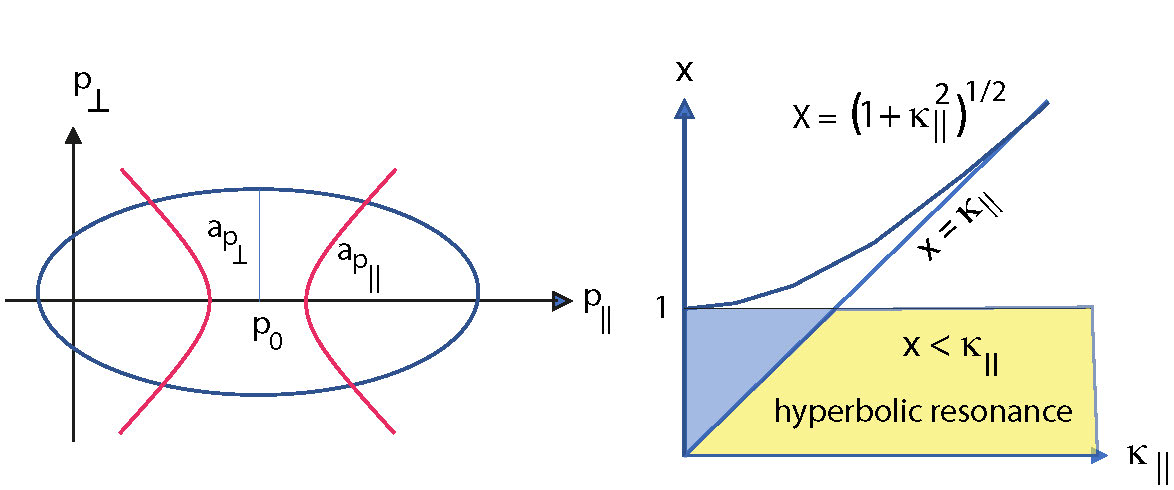}}
\caption{The different resonance topographies Eq. (\ref{eq-p-res2}) in dependence on the normalized frequency $x= \omega/\omega_{ce}$ in the fundamental band $x<1$. Elliptic resonance curves of the sort shown in the left part of the figure occur to the left of the line $x=\kappa_\|$ and below the curve $x^2=1+\kappa_\|^2$ in the range $x<1$ required by the fully relativistic resonance in the fundamental band $n=1$. Hyperbolic resonance is obtained in the yellow part on the right for $x<\kappa_\|$ and $x<1$. Resonant hyperboles are the kind indicated on the left in red colour. Apparently their range is substantially larger than that of the ellipses. But its extension depends on the restrictions on the large wave numbers that emerge from in the nonlinear evolution of the resonance. The range of resonance is in fact not restricted to proximity to $x=1$ as in the weakly relativistic resonance theory. At higher relativistic energy with $\gamma\gg1$,  frequencies well below $x=1$ become resonant. } \label{fig-hyp}
\end{figure}

Away from the local cyclotron frequency $\omega_{ce}$, however, growth is reduced on the X-mode branch where higher energy electrons $\gamma\gg0.1$ are required to be present in an excited state, a case barely existing in the auroral region. Here the group speed increases slightly before dropping to  low values when approaching the lower X-mode cut-off. In the auroral magnetosphere large numbers of relativistic electrons are absent  \citep[cf., however,][for a counter example]{xiao2016}. The relation between $k_\perp$ and $\gamma$ is
\begin{equation}
k_\perp^2\lambda_e^2+1\approx \frac{1}{\gamma^2}\frac{\omega_{ce}^2}{\omega_{e}^2}
\end{equation}
In the longer-wavelength regime $\lambda_\perp>\lambda_e$ one has from here 
\begin{equation}
1< \gamma \lesssim\frac{\omega_{ce}}{\omega_e}\big(1-{\textstyle\frac{1}{2}}k_\perp^2\lambda_e^2\big)
\end{equation}
 between $k_\perp$ and $\gamma$ away from the plasma resonance with resonant frequency reduced due to the increase in the required resonant energy. These wave numbers define the location of the resonance on the lower X-mode branch according to  (\ref{eq-disp-rel}).

The elliptical nature of the resonance condition is however not easily matched by any of the reasonable and measured electron distribution functions in the magnetosphere. Such a distribution should be some kind of a (phase space shifted)  \emph{hollow} beam or \emph{horseshoe} distribution with parallel shift $\beta_0$ (in terms of the momentum $p_0$),  highly diluted regions lacking  higher-energy electrons in the hole. Their boundaries are formed by accelerated/scattered cold electron beams \citep{muschietti1999,ergun1998}. This should provide the required positive $\beta_\perp$-space gradient mimicking an excited state of the resonant electrons. The beam-nature of the distribution is imposed by the finite displacement $\pm p_0$ along the external field which is seen in the observer's frame. This poses the question of the formation of such hollow beam distributions. Reasons for them can be found in the generation of electron holes by strong field-aligned currents and have been attributed to the downward current region in the auroral magnetosphere \citep{treumann2012} where such partial hollows have been assumed, either being electron holes or if of larger scale so-called horse-shoes.

Guided by auroral zone observations it however seemed, and from theory was also supported that the prevalent electron distribution of down going electrons, part of which is reflected, part absorbed by the dense ionosphere below, would become a loss-cone distribution. Numerical calculations based on the resonance ellipse and using observed loss cones \citep{omidi1984} found just very small amplification rates there. Moreover, from VLF observation and theory it became pretty clear that any loss cones are readily depleted or strongly reduced by intense interaction and amplification of VLF noise \citep{labelle2002}. The latter leaves little rudimentary resonance for the excitation of the ECMI and agrees well with the calculations based on the loss-cone. 

Other relevant electron source-distributions are caused in the generation of electron holes. These result in strong field aligned currents which decay into localized electrostatic structures of few Debye length $\ell_\|\sim \mathrm{few}\ \lambda_D$ extension along the magnetic field but large perpendicular scale $\ell_\perp\gg \ell_\|$ of the order of the perpendicular X mode wavelength $\ell_\perp\sim k_\perp^{-1}$. Inside those holes the conditions of excited electron states are satisfied. Here we do not go into a detailed investigation of this most interesting fact \citep{muschietti1999} to that we had referred in earlier work \citep{treumann2012}. In the next section we deal with the question which interests us here most: the condition of excitation of the ECMI on the lower X-mode branch and its possible escape into free space without requiring any usually evoked propagation effects in inhomogeneous plasmas \citep{zarka1986,louarn1990,lamy2010}.

Depending on the available electron energy and the excited phase space distribution function $f_e(p_\perp,p_\|)$, the ECMI will become effective over a large section of the lower-X mode branch to drive the X mode unstable. The range where this could happen is prescribed by the condition of a dilute plasma with $\omega_e^2\ll\omega_{ce}^2$ and the additional necessary condition on the distribution function that $\partial f_e/\partial p_\perp>0$ in an appropriate range on the resonance curve in the $(p_\|,p_\perp)$-plane. This resonance line for purely perpendicular propagation is given by
\begin{equation}
\sqrt{1+p_\perp^2/m^2c^2+p_\|^2/m^2c^2}=\omega_{ce}/\omega^l_X(k_\perp)
\end{equation}
with $\omega^l_X(k_\perp)$  the lower-X mode branch frequency with the right-hand side a function of $k_\perp$. Any of those unstable X-mode waves have of course frequency $\omega^l_X<\omega_{ce}$ below the nonrelativistic cyclotron frequency. They are confined to the plasma and detectable only \emph{in situ}, unless a mechanism is found which allows them to either escape into free space by tunnelling the ECMI-X-mode stop-band $\omega_{ce}<\omega<\omega_{co}^u$ between the cyclotron and upper X-mode cut-off frequencies. 

The other possibility is to undergo a three-wave interaction process as we proposed \citep{baumjohann2022} and will be investigated below. However, staying with strictly perpendicular propagation one may also envisage direct excitation of higher harmonics $n>1$ of the cyclotron frequency $\omega_X^u=n\omega_{ce}/\gamma$. Before returning  to the three wave process, we briefly discuss such

\subsection{Wave-wave interaction at the fundamental}
From observations is not entirely clear whether AKR is observed in the second harmonic or not. Fundamental radiation is trapped below the X-mode stop band (see Figure \ref{fig-disprel} )  and escapes only under strongly inhomogeneous conditions which have not yet been clarified convincingly. On the other hand, harmonic radiation when identified appears to be surprisingly intense. For the above noted reasons one may doubt in the importance of any direct radiation mechanism. The large growth rates $\Gamma_n\sim n\omega_{ce}$ respectively large amplification factors $\kappa_n$ they require are hard to reproduce in theory. Free space modes escape quickly before picking up any amplification unless the source extends over very large distances while maintaining all the conditions in favour of excitation. In addition to the obvious confinement of the fundamental $\omega_X\lesssim\omega_{ce}$ observation of apparently intense harmonic radiation \citep[cf., e.g.,][]{pottelette2001} in the spectrograms poses a problem. 

Of course, these conditions refer, in addition, to the required sufficient condition on the electron distribution function in phase space and are, probably, not very easy to satisfy in general. They require rather special electron distribution functions which, in order to contribute efficiently to growth,  must adapt to the particular geometry of the resonance curve in phase space. Such distribution functions must possess a positive perpendicular momentum gradient along the phase space resonance or at least large parts of it in order to pick up the contributions of as many electrons along the resonance as are elevated to the higher momentum/energy excitation level.  

Here we are interested in conditions which allow transformation of the ECMI at the lower X-mode branch into the free space mode. As argued above, at the lower X-mode branch the ECMI can indeed attain large amplification because of the comparably slow group velocity under the conditions below $\omega^l<\omega_{ce}$. Since waves  can be excited here in any direction perpendicular to the external magnetic field, a three wave process \citep{sagdeev1969,davidson1972} suggested in \citep{baumjohann2022}, becomes possible where two lower branch modes interact to compensate their large perpendicular wave numbers and result in a long wavelength mode at roughly the sum of the two frequencies according to the three-wave interaction conditions of conservation of total energy and momentum
\begin{equation}
\omega^u_X=\omega^l_{1X}+\omega^l_{2X}, \quad k^u_\perp=k^l_{1\perp}-k^l_{2\perp} \ll k^l_{1,2\perp}
\end{equation}
It generates a long wavelength $k_\perp^u\approx \omega_X^u/c$ upper branch X mode of frequency $\omega^u_X\lesssim 2\omega_{ce} $ below though near the second harmonic $n=2$. This wave is large amplitude, propagates in the free space mode, can escape without any difficulty from the source region and does not need any further amplification. The condition under that it may meet the free space branch is that its frequency must be above the upper X-mode-branch cut-off. Since each of the resonant waves on the lower branch satisfies the resonance condition $\omega_X^l=\omega_{ce}/\gamma$ the condition for escape of the ECMI-wave at the upper branch becomes
\begin{equation}
\frac{\gamma_1+\gamma_2}{\gamma_1\gamma_2}>1+\omega_e^2/\omega_{ce}^2
\end{equation}
where the indices refer to the two lower branch waves that participate in the interaction, and we neglect powers of small quantities. If the resonant energies do not differ much, this yields
\begin{equation}
1<\gamma <\frac{2}{1+\omega_e^2/\omega_{ce}^2}
\end{equation}
which of course is the same condition as for direct excitation below the second harmonic $n=2$ on the upper X mode branch given above. Its meaning in the present case is, however, quite  different. The two interacting waves are slow and have grown to large amplitudes. In head-on collisional interaction they result in large-amplitude escaping radiation just below the second harmonic $n<2$. For small frequency ratio $\omega_e/\omega_{ce}$ in the denominator, the marginal relativistic factor is close to $\gamma\lesssim2$ or, as already noted, the upper limit on the electron energy is $\epsilon_e\lesssim 0.9$ MeV. 

Low energies excite the ECMI on the lower X-mode branch close to $\omega_{ce}$, but higher energies near the upper energy limit do also contribute. They shift the resonance down on the lower branch according to $\omega^l_X=\omega_{ce}/\gamma$. Thus the ECMI can, in principle, become excited almost along the entire lower X-mode branch. Escaping radiation generated via the three-wave collision process is nevertheless allowed only for those waves excited by electrons of the above limited energies. Their energy determines the location of the resonance on the lower X-mode branch dispersion relation (\ref{eq-disp-rel}). 

We have $\omega^{2}\ll\omega_{co}^{u2}$ and neglect higher powers of $\omega_e/\omega_{ce}$, which is justified in all cases where the ECMI is expected to become effective. In that case the lower X-mode branch dispersion relation (\ref{eq-disp-rel}) simplifies yielding 
\begin{equation}\label{eq-disprel}
\frac{k_\perp^2c^2}{\omega_{ce}^2}\approx \frac{\omega^2-\omega_{co}^{l2}}{\omega_{ce}^2}=\frac{1}{\gamma^2}-\frac{\omega_e^2}{\omega_{ce}^2}
\end{equation}
for the  relation between $k_\perp$ and $\gamma$, which reproduces condition (\ref{eq-gamma-1}) while holding all along the lower X-mode branch
\begin{equation}
1<\gamma< \frac{\omega_{ce}/\omega_e}{\sqrt{k_\perp^2\lambda_e^2+1}}
\end{equation}
All wave numbers
\begin{equation}
k_\perp\lambda_e\lesssim\frac{\omega_{ce}}{\omega_e}\Big(1-{\textstyle\frac{1}{2}}\frac{\omega_e}{\omega_{ce}}\Big)
\end{equation}
are eligible for the three wave ECMI emission process above the upper cut-off frequency, including almost the entire lower branch wave numbers. 
Combination of the three last conditions on $\gamma$ and $k_\perp$ yields that the three-wave interaction will become effective for frequency ratios
\begin{equation}
\omega_{ce} > \sqrt{3}\ \omega_e
\end{equation}
a condition that is easily met in the auroral magnetosphere where under the conditions when emission of AKR occurs, the electron cyclotron frequency by far exceeds the plasma frequency.  

{To close this section, it is of substantial interest noting that exactly the same nonlinear wave-wave interaction mechanism of generation of radiation in the free-space X mode has been explicated in detail for the Z mode \citep{yoon2016} in the approximate second $n=2$ and fourth $n=4$ harmonics of the cyclotron frequency. This mechanism would in principle compete with the  ECMI wave-wave generation mechanism on the lower X mode branch proposed here, if not the two mechanisms would unfortunately exclude each other. The ECMI works solely under the condition that the ratio $\omega_e/\omega_{ce}\ll1$ is small. In this case the Z mode becomes confined to a rather narrow region around $x\approx \omega_e/\omega_{ce}$ \citep{melrose1986}. On the other hand, in the opposite situation when this ratio is large, as presumably for instance in the downward current region, then the ECMI becomes unimportant and excitation of the Z mode dominates (though see the discussion below in the section on downward currents). This is a very interesting case, indeed, as it predicts that in that case AKR could be radiated from nonlinear coalescence of  Z modes near the upper hybrid frequency $\omega\sim \omega_{uh}$. Generation of Z modes in that case is much less restricted than the ECMI on the X mode. It is well known that the Z mode is generated rather frequently by the continuous presence of VLF \citep{labelle2002} in the auroral magnetosphere where it is not necessarily found just in the auroral zone and at extreme magnetic activity but spreads over a  wider latitudinal and longitudinal region in space and in time. One expects that by the mechanism of wave-wave interaction in both, Z modes (possibly in the downward current region outside the upward current region) and X modes (preferentially in the upward current region), produce intense escaping AKR at nearly same harmonic frequencies in the free-space X mode.} 

\begin{figure}[t!]
\centerline{\includegraphics[width=0.75\textwidth,clip=]{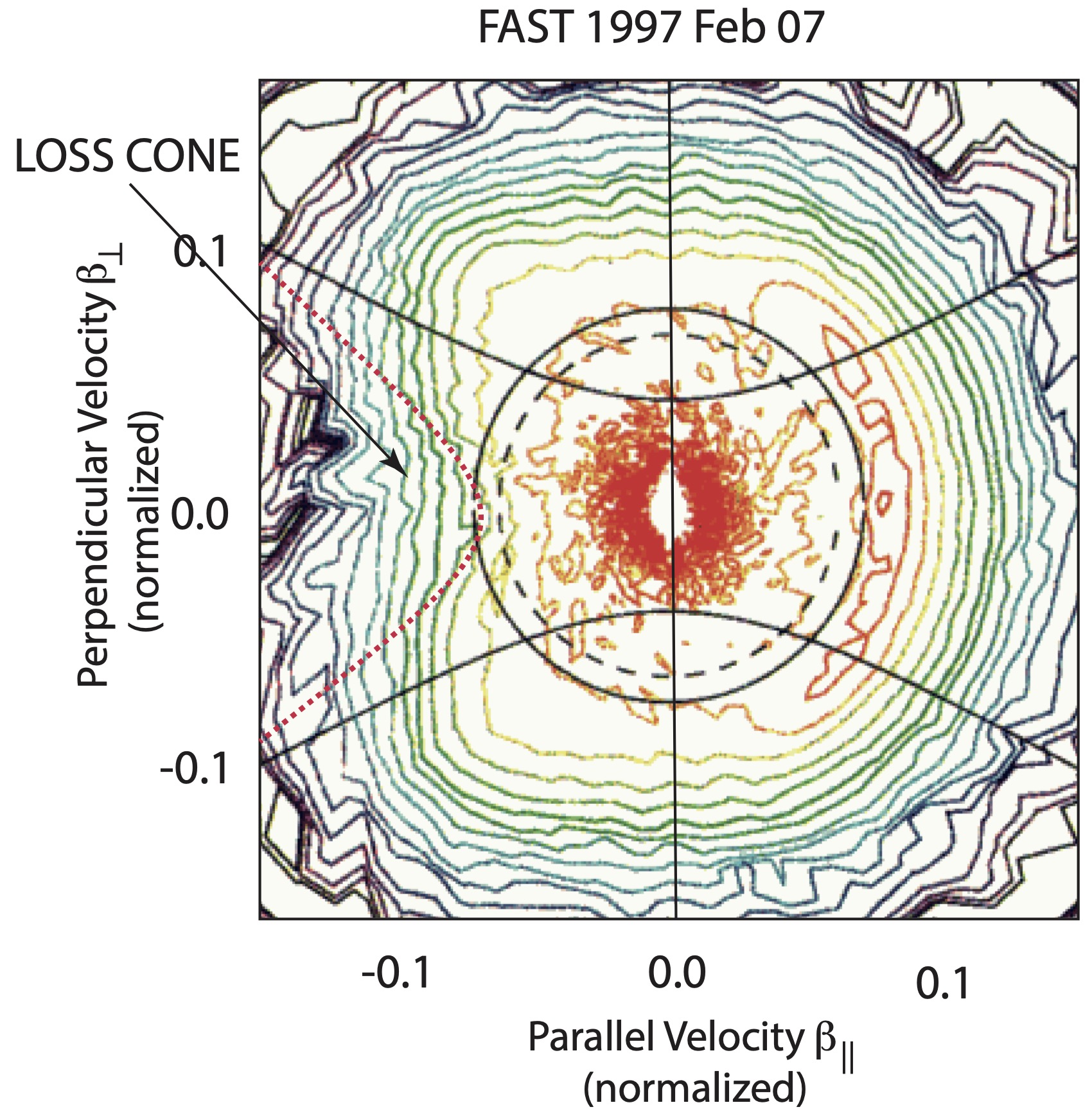}}
\caption{Measured loss-cone distribution in the presumable AKR upward current auroral source region \citep[after][which should be consulted for a detailed description of the observations]{delory1998}. The intensity of downward electron fluxes is colour-coded (from black to red covering 5 orders of flux magnitude) with red highest intensities, here indicating the denser warm auroral-magnetospheric electron background. The comparably wide upward (negative normalized parallel velocities $\beta_\|=v_\|/c<0$)  electron loss-cone is well exhibited. Circles show its theoretical (solid) and observed (dashed) low-speed bounds. At positive parallel speeds the red banana indicates the high energy  downward auroral electron beam. The two black hyperbolas are theoretical angular limits of inner and outer phase space regions. The dotted red hyperbola at negative (upward) $\beta_\|$ is the resonant hyperbola about along the loss-cone boundary with $k_\|c>\omega$. Positive perpendicular gradients along and inside this hyperbola contribute to resonant ECMI growth. The loss cone is wide enough to host a continuity of such hyperboles inside its bounds thus providing the ECMI a bandwidth $\Delta\omega$ at constant frequency $\omega$. Uncertainty of the observations is too large for calculating quantitative numbers.} \label{fig-delory}
\end{figure}

\subsection{Nonlinear evolution}
The ECMI-unstable lower X-mode branch beneath $\omega_X/\omega_{ce}\equiv x=1$ is a slow wave with comparably low phase velocity $v_{ph}=\omega/k<c$ and, in particular, low group speed $v_g=\partial\omega/\partial k<c$, both substantially less than the velocity of light. As argued in the previous section, these properties allow the wave to grow and achieve amplitudes which are large enough to violate the linear assumption. In this moderately or large amplitude state and being unable to leave the plasma below the X-mode stop band, the wave exerts a slowly variable ponderomotive force on the dilute low-pressure plasma background. (Note that the plasma is generally assumed to be at most temperate in this entire theory; what concerns the wave dispersion, so the background is even assumed to be cold.) The ponderomotive force results from an average over the entire ensemble of unstable waves thus being the effect of collective action. The variation of the ponderomotive force is slow; it occurs on the ion time scale. The plasma reacts to this force as described by the Sagdeev-Zakharov (slowly variable nonlinear Schr\"odinger) equation \citep[cf., e.g.][]{davidson1972,treumann1997} by exciting ion acoustic waves which under stationary conditions lead to envelope solitons, density depletions that trap the high frequency lower branch X-mode waves. The relative quasi-neutral density modulation $\delta N/N_0$ (amplitude of the ion-acoustic wave) of these  structures, which from pressure balance are density depletions (cavitons), is related to the X-mode amplitude as
\begin{equation}
\frac{\delta N}{N_0}\approx-\frac{\epsilon_0}{4m_iN_0c_{ia}^2}|\delta E_X|^2
\end{equation}
with $c_{ia}\ll c$ the ion acoustic speed. Trapping of the X-mode inside the density depletions causes a spatial modulation of the wave spectrum. (One may note that only the electric field amplitude appears here. The magnetic amplitude of the electromagnetic wave is relativistically small.) At the same time it leads to two other important effects.

Firstly, trapping of the spectrum of participating X modes splits the waves into two populations of oppositely directed wave numbers $\pm k_{tr}$ which bounce back and forth inside the density depletions. Several of these waves will thus participate in the above described three-wave process to cause second harmonic radiation $\omega_X\lesssim 2\omega_{ce}$. This  leads to  losses of wave energy and may ultimately terminate the deepening of the density depletions, i.e. stabilising the nonlinear ion-acoustic wave amplitude, even though radiative losses generally remain only weak.  

The second interesting effect concerns the trapped wave mode wavelength $\lambda$ (or wavenumber $k_X$). Conservation of caviton shape over the time of  caviton formation and evolution implies that the number of waves (wavelengths) inside the cavity is conserved and remains constant. Shrinking the cavity size thus necessarily shortens the wavelengths and increases the wavenumber $k_X$ according to
\begin{equation}
k^2\propto \epsilon_0|\delta E_X|^2
\end{equation}
with the right hand side the wave energy density which evolves during nonlinear formation of the caviton, i.e. the sum of the individual wave energies $\hbar\omega$ per volume of the caviton with frequency being conserved. This affects the dispersion of the trapped waves. This is very interesting by itself.

Though the trapped X mode waves inside the cavitons initially propagate in the linear X mode on the linear dispersion branch, shrinkage and shortening of wavelength modifies the linear X mode branch near resonance until it becomes nonlinear, a rather complicated process which strongly modifies the linear dispersion. In fact, due to the deviation of the dispersion curve from its linear topology that is caused by resonant interaction with the energetic source electron component, the X mode here becomes quasi-electrostatic which makes it active in affecting the plasma through its ponderomotive force and particle trapping inside the self-consistently generated cavitons \citep[cf., e.g.,][for the complete nonlinear theory of purely electrostatic waves]{schamel2022}. The resonance still occurs beneath though close to $\omega_{ce}$ here (unless a non-linear frequency shift is taken into account which is assumed of higher order, here, and thus negligible). Constancy of wave frequency implies that the nonlinear (about electrostatic) interaction stretches the dispersion curve out to large wave numbers $k_{nl}$ which substantially exceed the linear wavenumber $k$, while in frequency remaining below the electron cyclotron frequency thereby avoiding crossing it upward to approach the upper hybrid resonance $\omega_{uh}$ which is not allowed by the resonance condition (as long as it remains linear by itself). 

This increase in $k\to k_{nl}$ causes a severe additional  reduction of the phase velocity $v_\mathit{ph,nl}=\omega/k_{nl}\propto \omega/\sqrt{|E_X|^2}$. In addition the sudden self-consistent trapping of the waves inside the cavitons (resulting from the ponderomotive force) implies a violent retardation of the group speed to $v_g\sim c_{ia}$ which adjusts it to the ion acoustic velocity. The slow down in phase velocity substantially increases the spatial amplification rate, as has been argued above. As consequence, the waves locally experience many e-foldings and consequently also large wave amplification due to increase of their e-folding length. As a result of these two effects, the amplitude of the trapped X mode waves will readily grow to become large, which on the one hand causes additional deepening of the cavitons, while also strongly supporting wave-wave interaction and second harmonic radiation $\omega_X\lesssim 2\omega_{ce}$. Higher harmonic resonance, as explained above, has little chance to cause comparably intense radiation. Resonance $x=\omega/\omega_{ce}\lesssim1$ at the lower X mode branch in contrast amplifies the confined short-wave lengths X modes \citep[picked up, for instance, from thermal electromagnetic background noise, cf., e.g.,][]{yoon2017} which propagate at strongly reduced phase velocity and experience many e-foldings.

\section{Hyperbolic resonance}

The nonlinear deformation of the linear dispersion has another profound effect on the topology of resonance that so far {(at least to our knowledge)} has been missed in the literature on the ECMI. This is uncovered when considering the resonance condition (\ref{eq-p-res1}). The linear-state relativistic resonance is the famous ellipse in $\vec{\beta}$ or $\vec{p}$-space. In the nonlinear state, however, there arises the possibility for the ellipse to turn into a hyperbola
\begin{equation}\label{eq-hyperb}
\frac{(p_\|- p_0)^2}{a_{\|p}^2}-\frac{p_\perp^2}{a_{\perp p}^2}=1 
\end{equation}
This happens when from Eq. (\ref{eq-p-res2}) during nonlinear evolution and increasing wavenumber $\kappa=kc/\omega_{ce}$ the conserved normalized frequency $x=\omega/\omega_{ce}<\kappa_\|$ for some oblique propagation angle $\theta$ drops below the parallel normalized nonlinear wave number $\kappa_\|$. {(One may note that this leaves the numerators in (\ref{eq-p-res2}) invariant such that only the transverse radius $a_{\perp p}$ is affected to become imaginary).} The condition includes only the parallel wave number and is thus simply a condition on the the angle $\theta$ of propagation which means that
\begin{equation}\label{eq-hyp-cond}
\cos\theta>\omega/ck_{nl}=\frac{v_\mathit{ph,nl}}{c}
\end{equation}
exceeds the ratio of the strongly decreased non-linear phase velocity $v_\mathit{ph,nl}$ to light speed. {It says that for the topological switch to take place the angle of resonance should turn more parallel.} This condition is readily satisfied close to the gyrofrequency $\omega_{ce}$ already in the linear regime, as indicated in Figure \ref{fig-disprel} where the dispersion begins to flatten out shortly before crossing the line $x=1$. It essentially excludes perpendicular propagation from hyperbolic resonance, which however is clear anyway. The restriction of the resonance to $x<1$ in onsetting nonlinear evolution and the following smooth increase in wavenumber $k\to k_{nl}$ warrants that this condition is always met at oblique propagation satisfying the above condition which relaxes in further nonlinear evolution. {One may note that this important change in the resonance topology occurs when working in the fully relativistic resonance regime which applications have so far avoided. It is of particular importance in the astrophysically interesting ultra-relativistic regime.} 

Hence transition from  linear elliptic to nonlinear hyperbolic resonance is quite natural. In its course the resonating wave becomes increasingly oblique during nonlinear interaction. On the other hand, it is clear that this case cannot be realized for any of the resonances $n>1$ which are not confined and thus do not interact nonlinearly. 

 Manipulation of the relativistic resonance condition $x\gamma(\vec{p}) =1+\kappa_\|p_\|/m_ec$
 with constant $x\approx1$ yields an upper limit on 
\begin{equation}\label{eq-hyp-cond}
\cos\theta <\frac{\gamma-1}{\gamma\beta_\|}\frac{\omega_{ce}}{k_{nl}c}= \frac{\epsilon_\mathit{kin}}{m_ec^2\beta_\|}\frac{\omega_{ce}}{\omega_e}(k_{nl}\lambda_e)^{-1}
\end{equation}
where $k_{nl}$ is the increased nonlinear wavenumber. This is in fact no serious restriction, as long as the right-hand side exceeds unity. Otherwise it just excludes a range of nearly parallel propagation angles hence permitting  hyperbolic resonance over a wide oblique angular interval  $\Delta\theta$.

\subsection{Upward current region}
As it turns out, the upward current region becomes the ideal place for application of the hyperbolic resonance.  We shall demonstrate that it is best suited to fit the celebrated loss-cone distribution as primary source for generation of AKR here.

Observations in the presumable upward-current AKR source region of the auroral  magnetosphere \citep[e.g.,][]{pottelette2005} suggest that the kinetic energy of source electrons is about $\epsilon_\mathit{kin}\sim 10$ keV, while the frequency ratio amounts to about $\omega_{ce}/\omega_e\sim 10$. Parallel speeds of the auroral electrons have been measured roughly around $\beta_\|\sim 10^{-1}$. Moreover, the electron inertial length in the diluted upward current region is of order   $\lambda_e\sim0.5$ km. This yields within the uncertainty of these numbers
\begin{equation}
\omega/ck_{nl}<\cos\theta<1/k_{nl}\lambda_e 
\end{equation}
As the left-hand side readily holds close to $x=1$, there are no restrictions on the propagation angle here for any nonlinear wavelengths $\lambda_{nl}\gtrsim\lambda_e$ longer than the electron inertial length, a rather weak condition only. If $k_{nl}$ would further increase beyond this limit, the hyperbolic resonance becomes more oblique though still covering a large angular interval. 

It thus seems natural that the resonance in the upward current region on the lower X-mode branch, the presumably most important domain of its validity,  readily switches from elliptic to hyperbolic which naturally fits the loss-cone distribution and the interior of the loss-cone well if not much better than the resonant ellipse.   

This is shown in Figure \ref{fig-delory} on one example of available highest-resolution FAST observations of downward electron fluxes in the upward current region, performed two decades ago \citep{delory1998}. The figure suggests that, in contrast to the usually used resonance ellipse, the hyperbolic resonance conveniently covers the entire inner perpendicular electron flux (velocity space distribution) gradient $\partial f_e/\partial\beta_\perp>0$ offered by the interior of the loss cone for excitation of the ECMI lower-branch X-mode. This has some clarifying implications on the choice of the source-electron distribution function. Hyperbolic resonance is clearly in favour just of the celebrated loss-cone distribution as the main electronic phase-space source-distribution here where it is continuously observed and theoretically supported.  

The reasonable conclusion is that in the upward current region loss-cone distributions, like those in Figure \ref{fig-delory}  \citep{delory1998}, can under all circumstances drive the lower branch ECMI just beneath $\omega\lesssim\omega_{ce}$ (see Figure \ref{fig-disprel}) increasingly unstable and entering its state of nonlinear evolution towards large wavenumbers $k\lambda_e<1$. This is expected to happen already for rather moderate amplitudes readily deforming the linear dispersion relation into its nonlinear cousin to participate in hyperbolic rather than elliptic resonance. 

 In the dilute upward-current low-pressure plasma background, it should be stressed that, even rather moderate confined ECMI-X mode amplitudes suffice to enter the nonlinear state by choosing a  suitable initial wavenumber and frequency interval that the thermal electromagnetic background noise offers for amplification \citep[for the presence of such noise, which for oblique propagation exists in particular  below and close to $\omega_{ce}$ see,][]{yoon2017}.

Such a switch in resonance suites the upward loss-cone distribution quite well for a spectrum $\Delta k_\mathit{res}$ of wavenumbers $k$ and excitation of a narrow spectral band $\Delta\omega$ below $\omega_{ce}$. These intervals are determined by the angular width of the loss cone and the steepness of the resonant perpendicular momentum-space gradient. 

As described in the  caption of Figure \ref{fig-delory}, the circles separate the loss cone from the unaffected main plasma distribution at low electron speeds. The observations do not allow for unambiguous identification of a parallel shift $p_0$ of the distribution. The two drawn black hyperbolic lines map the boundaries of the loss-cone in their upward directed parts while being artefacts on the downhill side. Tentatively one resonant hyperbola (red) has been drawn along the boundary of the loss-cone. The entire positive perpendicular momentum/velocity gradient region in the loss-cone could be filled with a continuum of such resonant hyperboles. At any fixed resonant parallel momentum $p_\|$ respectively velocity $\beta_\|$ this determines the growth rate $\Gamma_{n=1}(p_\|,\omega)$ as function of resonant frequency and provides an estimate of the bandwidth of excitation which observations in situ the AKR source suggest to be of the order of just few kHz in so-called elementary events \citep[cf. the discussion in][]{baumjohann2022}. Unfortunately, this is inhibited by the large scatter of data in the iso-flux lines. 

Radiation in the fundamental X-mode is confined and cannot escape, at least not locally, unless it becomes scattered and propagates up along though oblique to the magnetic field until becoming gradually transformed to find itself on the free-space X-R mode branch in the magnetospheric tail. Otherwise the large-amplitude X mode remains trapped, as described above, it undergoes nonlinear wave-wave interaction to generate escaping second harmonic radiation. Presumably this is the reason for observation of locally intense though (because of their proximity to $x=1$) very narrow-band plasma-confined ECMI emissions near the fundamental $x\sim 1$ \citep{pottelette2005,baumjohann2022} in the upward current region. Wave-wave interaction of those confined large amplitude amplified waves transforms the confined X mode into the free-space escaping X-mode radiation at frequency $\omega\lesssim 2\omega_{ce}$ below the second harmonic.

\subsection{Downward current region}
The same hyperbolic resonance condition (\ref{eq-hyp-cond}) holds of course also in the downward current region,
but the mostly confirmed and reasonable absence of any loss-cone distribution here demands a different mechanism to cause positive perpendicular phase space gradients on the source-electron distribution function along the resonance hyperbole. Otherwise, if no appropriate electron distribution becomes available in the downward current region favourable for hyperbolic resonance, one requires $v_{ph}\sim kc$, in which case the resonance returns to elliptic. It seems that this applies indeed to the downward current region even though this condition is difficult to satisfy. It implies that the lower-branch X-mode waves maintain comparably large phase velocities and the resonance is replaced away from the electron cyclotron frequency, a condition difficult to achieve.

Unfortunately no comparably high-resolution measurements of the angular electron fluxes and distribution functions are yet (at least to our knowledge) available here. The precise form of the electron distributions is thus not known and would indeed be worth to be focused on in (hopefully planned and available) future auroral magnetospheric space missions in the spatially rather narrowly extended auroral downward current regions of which during substorms there are many in a row. Electron fluxes are upward, low energy, spatially highly structured and highly variable in time. The corresponding upward currents are not smoothly distributed over a wide latitudinal interval as in the upward current region where the comparable smoothness of the upward sheet current is reflected in the smooth spatial course of the current-transverse magnetic field component that is typical for a broad about unstructured main-field-aligned sheet current. 

In contrast to the upward current region, the downward main-field-aligned currents flow in latitudinally-narrow parallel current sheets or current braids which are kept apart by the comparably very strong external geomagnetic field. Accordingly the current-transverse magnetic field component fluctuates considerably spatially, typical for the presence of many narrow current filaments. The main geomagnetic field is strong enough to compensate the Lorentz force of these quasi-stationary narrow field-aligned sheet-currents which attract each other but cannot merge. Located at the boundary of the auroral cavity, average densities in the downward current region are comparably high though still in the range $\omega_e/\omega_{ce}<1$  \citep{temerin1998}. Energies of the upgoing current-carrying current-closing electrons of ionospheric origin are just around $\epsilon_\mathit{kin}\sim$ few keV at most, roughly one order of magnitude less than in the upward current region. Related upward velocities $\beta_\|$ decrease by a factor of roughly ten, about compensating the decrease in kinetic energy. 

The downward currents are intense enough for causing current instability, including reconnection in strong current-parallel guide fields and, in particular, nonlinear evolution \citep{carlson1998} which structures and deforms the electron distribution function and generates large numbers of (so-called) Debye-scale electrostatic structures \citep{bernstein1957,ergun1998}, ion and electron holes that propagate along the magnetic field. Debye lengths are the order of $\lambda_D\lesssim$ 10 m here, depending on the exact spatially variable density and temperature. Whether this is in favour or not of the ECMI remains unclarified. Absence of loss-cones makes the hyperbolic resonance less attractive requiring rather particular source-electron phase-space distributions. These may be provided by those nonlinear Debye-scale structures, current-driven electron holes. It has been suggested \citep[][and others]{muschietti1999} that holes cause violent deformations of the electron distribution function digging phase-space holes into it which contain a highly diluted temperate low pressure electron population and are bound by denser cold walls in phase space of much higher hole-field accelerated speed. At these boundaries the phase-space distribution develops steep gradients which may serve the needs for exciting the ECMI. This model has been used to propose the action of the ECMI here as well, based on elliptic, not hyperbolic resonance \citep{treumann2011,treumann2012} in those incomplete hollow-electron phase space distributions, sometimes called horse-shoes.

Observed electron holes extend several $\lambda_D$ along the magnetic field but are much less restricted in perpendicular extension, which is limited by the width of the unstable field-aligned current filaments and electron gyroradii, the latter being the order of $r_{ce}= \beta_\perp c/\omega_{ce}\lesssim 1$ km for the dominant source electrons. Trapping of linearly excited X modes in these (for the purpose of the ECMI pre-existing) plasma depletions \citep[which result in the nonlinear evolution of current instability, for the most recent complete theory cf.][]{schamel2022} is preferably at $k_\perp> k_\|$, i.e. short perpendicular wavelengths and slightly oblique propagation $\theta\lesssim90^\circ$. This is supported close to $x=\omega_X/\omega_{ce}\sim 1$ on the lower-branch X mode where wave numbers increase and wave group speeds become low, holding however only for the perpendicular wavenumber. This means that $\kappa_\|<x$ is a reasonable option inside those holes, whose Debye-structured property applies only to the parallel direction. Consequently, the electron-hole trapping-hypothesis identifies the resonance with an ellipse instead of the hyperbole in the adjacent upward current region. 

The range of X mode frequencies fits these perpendicular wavelengths which become trapped in parallel Debye-scale holes. The elliptic resonance condition, applied to  the interior of such oblate holes, supports excitation of the ECMI. In contrast to the upward current region, ECMI is caused at the perpendicular boundary of the phase space holes, not the general form of the meso-scale electron distribution as this, here, is not of the loss-cone family. Rather it is self-consistently provided by the self-consistent nonlinear evolution of the electron holes. Generation of electron holes and related phase space distributions of the hollow/horseshoe type is primarily independent on the excitation of the X mode, being provided by the nonlinear  evolution of sufficiently strong field-aligned electron currents \citep[as for electrostatic waves has extensively been reviewed by][]{schamel2022}. It is the particular electron phase space distribution resulting from this nonlinear evolution which, in the downstream current region, may encourage the excitation of lower-branch X modes in elliptical or hyperbolic resonance and support their increase to reach large local amplitudes while also experience wave-wave interaction to escape into free space.

The lower-branch hole-trapped X mode has large parallel wavelengths $\lambda_\|\gg\lambda_D$ along the magnetic field, which by far exceeds the Debye-scale. It overlaps a large number of Debye-scale low density holes which are densely chained along the magnetic field effectively experiencing  an  amplification from all those holes along the field over its parallel wavelength. Simultaneously the wave bouncing back and forth synchronously inside all of them in perpendicular direction. Even though each hole contributes differently to the growth of the wave, the concerted action of all holes in the average should lead to a substantial amplification of the X mode. If this is the case one expects that observations in situ will detect highly temporarily structured narrow-band and intense confined short perpendicular wavelength X mode radiation which propagates at group velocity substantially below light speed. Its frequency will be very close to the local electron cyclotron frequency $\omega\sim\omega_{ce}$. Growing to large amplitude this slow radiation will again undergo wave-wave interaction to generate second harmonic radiation which escapes from the source and can be observed from remote. The trapped large amplitude fundamental narrow-band X-mode radiation can, however, be observed only in situ \citep[as for instance in Figure 2 of][]{baumjohann2022} where it moves together with the entire chain of holes along the magnetic field. If remaining trapped for all the life-time of the holes, it will \citep[as has surprisingly been observed, see][and references therein]{labelle2022} also be transported away from the source by the holes either down to the ionosphere \citep{treumann2012a} or upward into the magnetosphere. Similar observation in the upper ionosphere \citep{parrot2012,parrot2022} have also been reported from DEMETER spacecraft observations.

\section{Conclusions}
{Examination of the fully relativistic ECMI  resonance condition in application to the auroral magnetosphere reveals a so far missed domain where the elliptic resonance  turns into a hyperbolic resonance therefore becoming topologically completely different matching the observation of loss-cones in the auroral upward current region and suggesting a number of interesting facts. Firstly, loss cone distributions with harrow and partially filled loss cones can nevertheless in fully relativistic theory become efficient sources of rather narrow band excnitation of lower branch X modes which cannot escape from the source region. The bandwidth in frequency of those modes would be determined by the narrow loss cone  boundary at given wave number which by observation is of the order of about at most few kHz. It can be observed only in situ the ECMI source. Second, these locally excited waves have low phase and group velocities, are trapped in the source region, experience several exponentiations, and evolve up to large amplitudes. This generates nonlinearities and causes the wave numbers to shrink further until the nonlinear wave to become trapped in cavities where the wave undergoes wave-wave interaction to generate second harmonic radiation above the upper X-mode cut-off on the free space mode. From there the radiation escapes. This picture rounds up the theory of AKR excitation by the ECMI in the very low density upward current region. } 

{Whether the ECMI can also work in the downward current region, remains uncertain. Primarily the condition of very low density is not necessarily given there. In that case the ECMI should not evolve. However, the downward current region is the location of very many so-called Debye-scale structures, very low density electron holes which along the field are of length of tens of Debye lengths but in perpendicular direction extend over many electron gyro-radii  corresponding to the transverse wavelength of the X mode. In perpendicular direction the X mode fits the hole and, in the low density region of the holes, becomes as well amplified and evolves nonlinearly. ECMI generated AKR could thus also result from here though in completely different ways.}

{In addition, however, in the downward current region Z mode radiation may compete in all those regions which do not experience dilution by Debye structures. Here the Z mode propagates and is excited by VLF, undergoes wave wave interaction and radiates at second and possibly even higher harmonics. This mechanism is not based on the ECMI but deserves to be taken into account when dealing with the downward current region.}  

{Another most interesting observation of AKR is that it seems to weakly leak down into the ionosphere at spacecraft altitudes and even down to the ground under favourable though still unknown conditions. These observations pose the interesting question: How can AKR generated at altitudes of at least 1000 km above ground pass into and even across the ionospheric density screen? Such a passage seems forbidden. However, if generated in density cavities, either self-generated cavities in the upward or within Debye scale cavities in the downward current region, then X modes could possibly be trapped inside those cavities and together with them may become transported down to the ionosphere where few of them could survive the transport and would leak out.  This most interesting observation still awaits its theoretical explanation.}

\begin{acknowledgement}
This work was part of a brief Visiting Scientist Programme (of RT) at the International Space Science Institute Bern. We acknowledge the interest of the ISSI directorate as well as the hospitality of the ISSI staff, in particular the assistance of the librarians Andrea Fischer and Irmela Schweitzer, and the Systems Administrator Willi W\"afler. 
\end{acknowledgement}




\end{document}